\documentclass[useAMS,usenatbib]{mn2e}
\usepackage[pdftex]{graphicx}
\usepackage{amsmath,amssymb}
\def\rmscr#1{{\hbox{\rm \scriptsize #1}}}

\def\msun{{\rm M}_\odot}

\begin{document}
\title[Diffractive Microlensing II]{Diffractive Microlensing II:
  Substellar Disk and Halo Objects}
\date{Accepted 2010 October 4.  Received 2010 September 29; in
  original form 2010 February 16}
\author[J. S. Heyl]{Jeremy S. Heyl\\
Department of Physics and Astronomy, University of British Columbia,
Vancouver, British Columbia, Canada, V6T 1Z1;\\
 Email: heyl@phas.ubc.ca; Canada Research Chair}
\pagerange{\pageref{firstpage}--\pageref{lastpage}} \pubyear{2009}

\maketitle

\label{firstpage}

\begin{abstract}
  Microlensing is generally studied in the geometric optics limit.
  However, diffraction may be important when nearby substellar objects
  lens occult distant stars.  In particular the effects of diffraction
  become more important as the wavelength of the observation
  increases.  Typically if the wavelength of the observation is
  comparable to the Schwarzschild radius of lensing object,
  diffraction leaves an observable imprint on the lensing signature.
  The commissioning of the Square Kilometre Array (SKA) over the next
  decade begs the question of whether it will become possible to
  follow up lensing events with radio observations because the SKA may
  have sufficient sensitivity to detect the typical sources, giant
  stars in the bulge.  The detection of diffractive lensing in a
  lensing event would place unique constraints on the mass of the lens
  and its distance.  In particular it would distinguish rapidly moving
  stellar mass lenses (e.g. neutron stars) from slowly moving
  substellar objects such freely floating planets.  An analysis of the
  sensitivity of the SKA along with new simple closed-form estimates
  of the expected signal applied to local exemplars for stellar radio
  emission reveals that this effect can nearly be detected with the
  SKA.  If the radio emission from bulge giants is stronger than
  expected, the SKA could detect the diffractive microlensing
  signature from Earth-like interstellar planets in the solar
  neighborhood.
\end{abstract}

\section{Introduction}
\label{sec:introduction}

Gravitational microlensing surveys discover microlensing events at an
increasingly rapid pace as the sensitivity of the instruments improve
and the observing strategies are optimised.  As the number of events
increase, about ten percent of the events have Einstein-diameter
crossing times shorter than a week and about one in fifty events have
times less than four days.  \citet{Stefano:2009p1904} has argued from
the dynamics of the Galaxy that these short-duration events are either
planet mass objects (freely floating or bound to a star), rapidly
moving stellar-mass objects such as pulsars (hypervelocity stars) or
events with small separations between the lens and source.  The third
possibility can be estimated statistically.  Furthermore Di~Stefano has
argued that the second possibility is much more likely than the first
due the rarity of high-velocity objects, and regardless of the type of
lens extensive follow-up of short-duration microlensing events is
warranted.

The first cut to determine the identity of the lens would be to look for
finite-source effects that could indicate that the angular size of
Einstein radius is not much larger than the source.  This would point
to a planet orbiting a star or a hypervelocity star.  Without
finite-source effects and without a positive identification of the
lens with a parent star or a neutron star, it is difficult to
distinguish a freely floating planet from a neutron star, for example.

This paper outlines a simple technique to distinguish these
possibilities and probe the presence of freely floating planets in the
solar neighborhood. The Square Kilometre Array (SKA) is planned to be
the most sensitive radio telescope in the foreseenable future
\citep{Schi07}, fifty times more sensitive than those existing
today. It provides a great benchmark for the observability of
diffractive microlensing.  By observing the lensing event at radio
wavelengths at the SKA, the diffractive lensing comes into play for
substellar objects but not for stars, so a comparison of the radio and
optical light curve could provide a direct measurement of the mass of
the lens.

The first section, \S\ref{sec:calculations}, outlines microlensing in
the diffractive regime and specifically examines the size of the
fringes in the short-wavelength limit (\S\ref{sec:short-wavel-limit})
to quantify the finite-source effects on observations of the fringes.
A particular equation of state for substellar objects
(\S\ref{sec:equation-state},\S\ref{sec:regimes}) defines the regime
in which lensing is important \citep[versus occultation,
see][]{2009arXiv0910.3922H}.  The letter outlines new
closed-form estimates for the variation in the light curves
(\S\ref{sec:light-curves}) (\S\ref{sec:results}), the sensitivity of
the SKA to the variation (\S\ref{sec:detection}) and the possible sources
(\S\ref{sec:sources}).  Finally \S\ref{sec:conclusions} places the
various possible sources in context and argues what the most promising
sources are.

\section{Calculations}
\label{sec:calculations}

\citet{1992grle.book.....S}  give the magnification for a point source
including diffraction
\begin{equation}
\mu_\omega =  
\left | \int_0^\infty u^{1-if} e^{iu^2/2} J_0(uv) du \right |^2.\label{eq:1}
\end{equation}
where $u$ is the radial coordinate that integrates over the plane of
the lens and $v$ is the impact parameter of the source relative to the
lens.  Both $u$ and $v$ are dimensionless and measure lengths in
units of the reduced Fresnel length,
\begin{equation}
l_\rmscr{Fr} = \sqrt{\frac{c}{\omega_d} \frac{D_d
    D_{ds}}{D_s}}
\end{equation}
Hence the value of $u_d$ which compares the angular size ($r_d/D_d$) of
the occulting portion of the lens to angular scale of its diffraction
pattern ($\lambda/r_d$) is given by
\begin{equation}
u_d = r_d \sqrt{\frac{\omega_d}{c} \frac{D_s}{D_d
    D_{ds}}} = \frac{r_d}{l_\rmscr{Fr}}. \label{eq:6}
\end{equation}
The limit where the gravitational field of the lens is negligible is
$f=0$, so the effect of gravity on the form of the integral is quite
modest.  The parameter $f = 2 R_S \omega_d/c$ ($R_S=2 G M_d/c^2$)
where $M_d$ is the mass of the lens and $\omega_d$ is the frequency of
the radiation at the lens.  The Einstein radius is the characteristic
length of the lens,
\begin{equation}
R_E = \sqrt{2 R_S \frac{D_d D_{ds}}{D_s} } = \sqrt{f} l_\rmscr{Fr} \label{eq:2}
\end{equation}
for the Schwarzschild lens where $D_s$ is the distance to the source,
$D_d$ is the distance to the lens and $D_{ds}$ is the distance between
the source and the lens.

The integral can be calculated in closed form in terms of the
confluent hypergeometric function ($_1F_1(a;b;z)$). In general, using
relation (6.631.1) in \citet{Grad94} gives the following
result 
\begin{eqnarray} \int_0^\infty u^{1-if} e^{iu^2/2} J_0(uv) du
  \!\!&=&\!\!e^{\pi f/4} e^{i \left ( \pi - f \ln 2\right)/2}
  \Gamma \left (1 - i \frac{f}{2}\right ) \times \nonumber \\
  & & ~{}_1F_1 \left ( 1 - i \frac{f}{2}; 1 ; -i\frac{v^2}{2} \right
  )\label{eq:3}.
\end{eqnarray}
The result for $f=0$ is simply $i\exp(-iv^2/2)$.  

\subsection{Short-wavelength limit}
\label{sec:short-wavel-limit}

It is very useful to look at the magnification in the limit of large
$f$ and small $v$.  Using the properties of the confluent hypergeometric
function or equivalently the WKB approximation
\citep{Deguchi:1986p1910} gives the following expression for the magnification
\begin{equation}
\mu_\omega \approx \pi f \left [ J_0\left ( \frac{f v}{R_E} \right ) \right]^2.
\approx \frac{R_E}{v} \cos^2 \left ( \frac{f v}{R_E} - \frac{\pi}{4}
\right ) 
\label{eq:4}
\end{equation}
where the second approximation holds for $f v/R_E > 1$
\citep{Deguchi:1986p1911}.  
This expression gives a useful estimate of
the angular size of the interference fringes on the sky in particular,
\begin{equation}
\Delta \theta_\mathrm{fr} \approx \frac{\pi}{f} \frac{R_E}{D_d},
\label{eq:5}
\end{equation}
so for $f>\pi$ or $\lambda < 4 R_S$ the criterion for the source to be
point-like is more stringent for the fringes than for the calculation
of the magnification assuming geometric optics.
\citet{1991ApJ...374L...5P} also look at the short-wavelength limit
but look at the path-length difference along the two geodesic paths to
get an alternative expression.

\subsection{Equation of State}
\label{sec:equation-state}

To determine whether occultation may be important, in particular
whether the object is larger than its Einstein radius, a mass-radius
relation or equation of state is required.  \citet{Chabrier:2008p1898}
provide a mass-radius relationship for giant planets through to
low-mass stars as a function of age.  Choosing an age of 5 Gyr fixes
the mass-radius relation above a mass thirty percent greater than that
of Jupiter.  Below this mass the observed masses of Earth, Uranus,
Saturn and Jupiter provide a mass-radius relation.  This mass-radius
relation is simply used to delineate the region within which occultation could
play a role as well as lensing 
\cite[see][for further details]{2009arXiv0910.3922H}.  The current
work focuses on the light curves for pure lensing (well above the green
line in Fig.~\ref{fig:regimes}).

\section{Results}
\label{sec:results}

\subsection{Regimes}
\label{sec:regimes}

\begin{figure}
\includegraphics[width=3.4in]{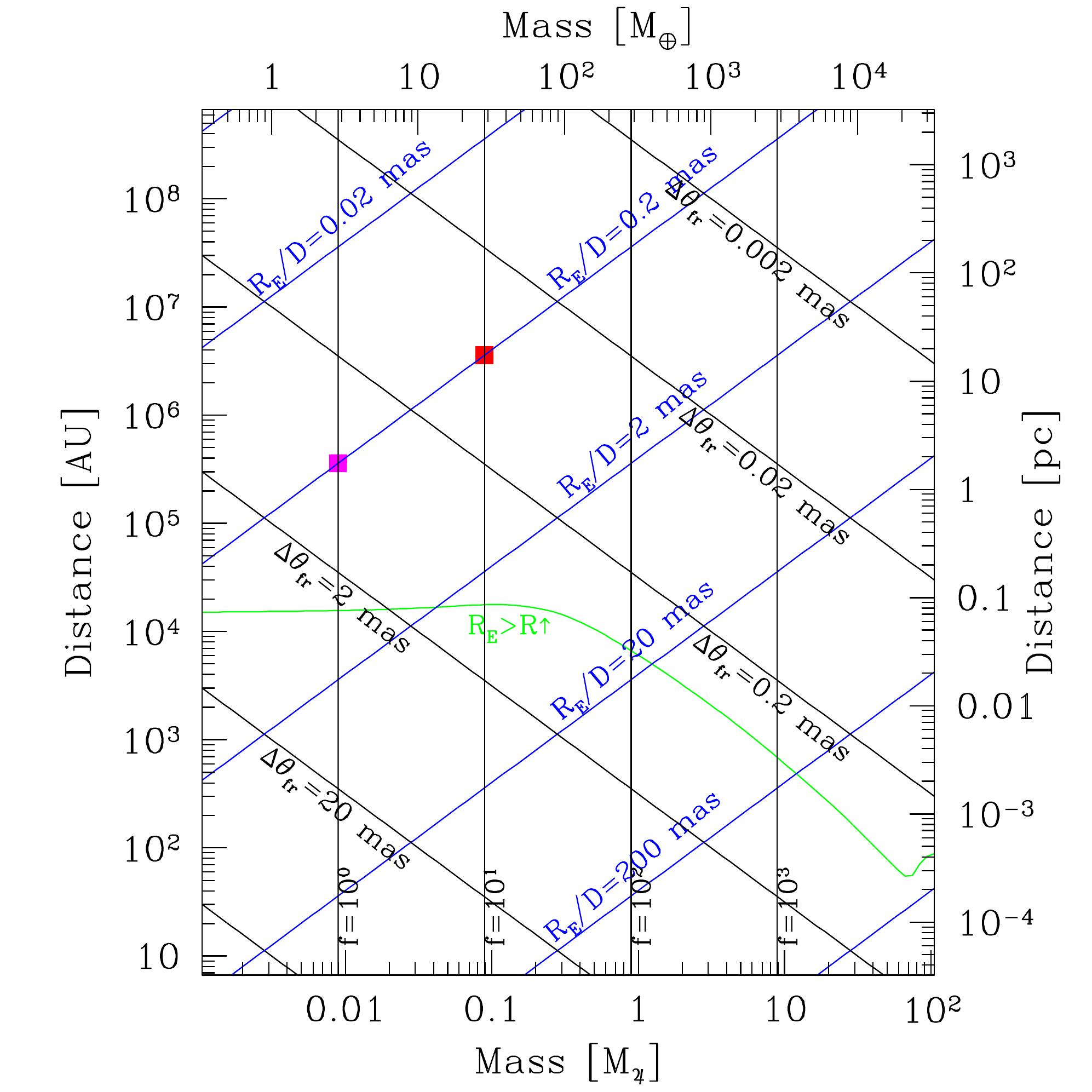}
\caption{The values of $f$, $R_E$ and $\Delta \theta_\mathrm{fr}$ as a
  function of the mass of the substellar object and its distance for
  $\lambda=10^{1.5} {\rm cm}$ or $\nu\approx 1{\rm GHz}$.  The value
  of $f$ scales as $\nu$ or $\lambda^{-1}$.  The slanted black lines follow
  the locus of constant fringe angular size in the limit of large $f$.
  They move upwards in proportional to the wavelength. The green curve
  $R_E>R$ indicates the region within which the Einstein radius is larger
  than the radius of the object so lensing will dominate over
  occultation, assuming the equation of state of
  \citet{Chabrier:2008p1898} above the mass of Jupiter and the
  mass-radius relation defined by Saturn, Uranus and Earth for lower
  masses.  Lines indicating both the angular size of the
  Einstein radius and the value of $f$ are shown as well.}
 \label{fig:regimes}
\end{figure}
The region of the solar neighborhood and mass of the intervening
substellar object where lensing is important is shown in
Fig.~\ref{fig:regimes}.  In particular above the line $R_E>R$ lensing
begins to dominate over occultation.  For the masses of interest this
corresponds to objects farther than about a tenth of a parsec;
therefore, beyond a parsec lensing strongly dominates.  If the source
is a giant in the bulge its angular radius is typically about 
20~$\mu$as.
More distant sources will appear smaller; however, they would be
difficult if not impossible to detect with the SKA.  If the lens lies
well below the blue line ($R_E/D=0.02$~mas) and the source is a bulge
giant, then the optical light curve would be similar to that of a
point source.  If the lens lies above or near the black line ($\pi
R_E/(f D)=0.02$~mas), the fringes would begin to be washed out due to
the finite size of the source; therefore, if a lensing event on a
bulge star lacks finite-size effects in the optical and lacks fringes
at one-metre, then the lens must lie the wedge well above the black line
and well below the blue line.  One could look for fringes at lower
frequencies to further restrict the properties of the lens; however,
the sources might be difficult to detect at such low frequencies
(\S\ref{sec:sources}).

\subsection{Light Curves}
\label{sec:light-curves}

\begin{figure}
\includegraphics[width=3.4in]{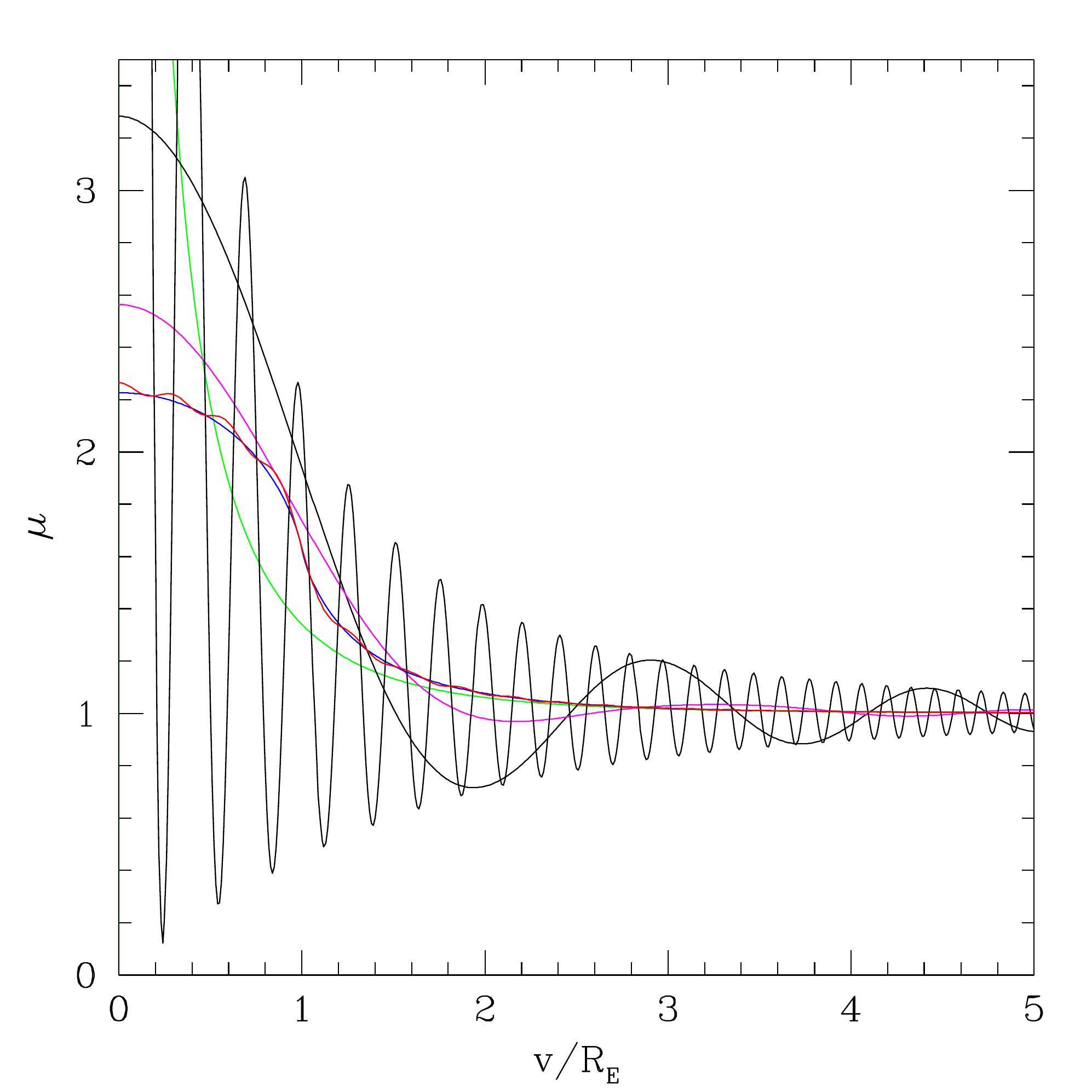}
\caption{The values of the magnification $\mu$ as a function of the
  distance between the source and the centre of the lens in units of
  the Einstein radius, $v/R_E$.  The result for $f=1$ is the slowly
  varying sinusoidal curve and $f=10$ is the more rapidly varying
  one.  The magnification from geometric optics is plotted in green.
  Notice for $f=10$ there are about three peaks over a length of one
  Einstein radius.  The other colours assume that the angular radius of the
  source equals the Einstein radius.  Blue is the geometric optics
  result, red is for $f=10$ and magenta is $f=1$ (see also the squares
  in Fig.~\ref{fig:regimes}).
}
 \label{fig:sample}
\end{figure}

More tantalizing than a null result would be the detection of fringes
(Fig.~\ref{fig:sample}) or a lack of magnification at the longer
wavelength.  By comparing the optical light curve (green in
Fig~\ref{fig:sample}) to the light curve at lower frequencies one
could determine the value of $f$ for the lens at the lower frequency.
Even if the fringes are diluted by finite source effects, the
diffraction does still cause the light curve to vary.  Typically if
there are six fringes across the disk of the source (red curve), the
difference between the light curve according to geometric optics and
the diffractive result is a few percent.  For twenty fringes over the
stellar disk, the difference is a few parts per thousand.  If the
source is a bulge giant (with a radius twenty times that of the sun),
this would correspond to the uppermost black line labelled
``$\Delta\theta_\mathrm{fr}=0.002$mas.''

For values of $f<1$ there is typically little magnification.  The
absence of magnification at a particular wavelength would restrict the
lens to have a Schwarzschild radius less than that wavelength --- a low
mass lens.  On the other hand, the detection of some magnification
along with fringes would determine the value of $f$ constraining the
lens to lie on one of the vertical lines in Fig.~\ref{fig:regimes} and
giving the mass of the lens.  The distance to the lens would also be
constrained by the relative amplitude of the fringing if the angular
size of the source is comparable or larger than that of fringes.

\begin{figure}
\includegraphics[width=3.4in]{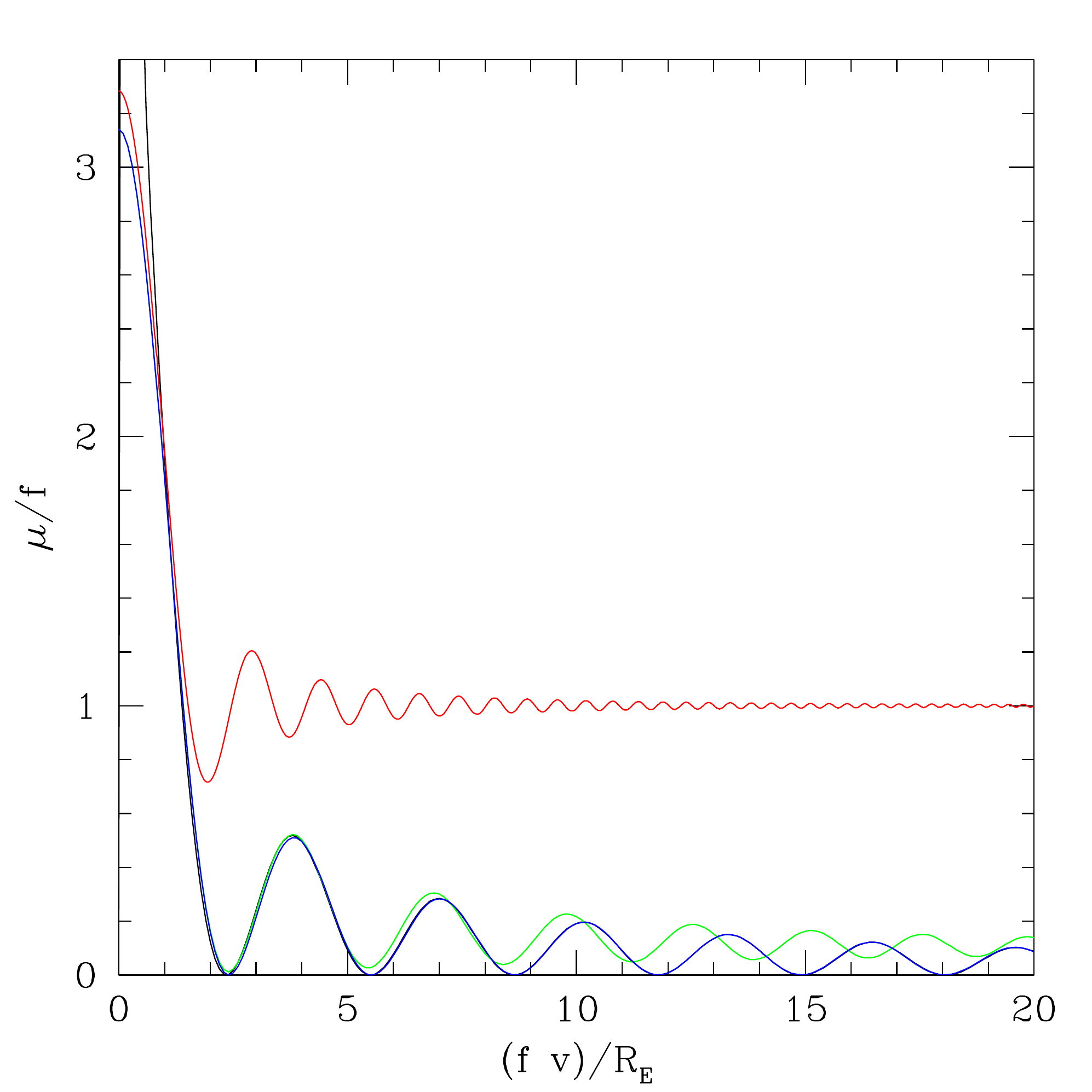}
\caption{Red, green and blue curves trace the magnification
  as a function of distance from the centre of the lens for $f=1,
  10$ and 100 respectively.  The magnification
  according to Bessel function in Eq.~\ref{eq:4} is 
  indistinguishable from the blue curve ($f=100$). The black curve
  gives the approximation using the cosine in Eq.~\ref{eq:4}.}
 \label{fig:fringe}
\end{figure}

For wavelengths less than a metre and lens masses greater than a tenth
of a Jupiter mass, the value of $f$ is much greater than unity.  In
this limit the fringe pattern approaches a simple cosine dependence
(Eq.~\ref{eq:4}) as can be seen in Fig.~\ref{fig:fringe}.  In the
limit of large $f$ the peak magnification is simply $\pi f$ while in
the geometric optics limit the magnification diverges as the source
lens and observer become aligned.

\begin{figure}
\includegraphics[width=3.4in]{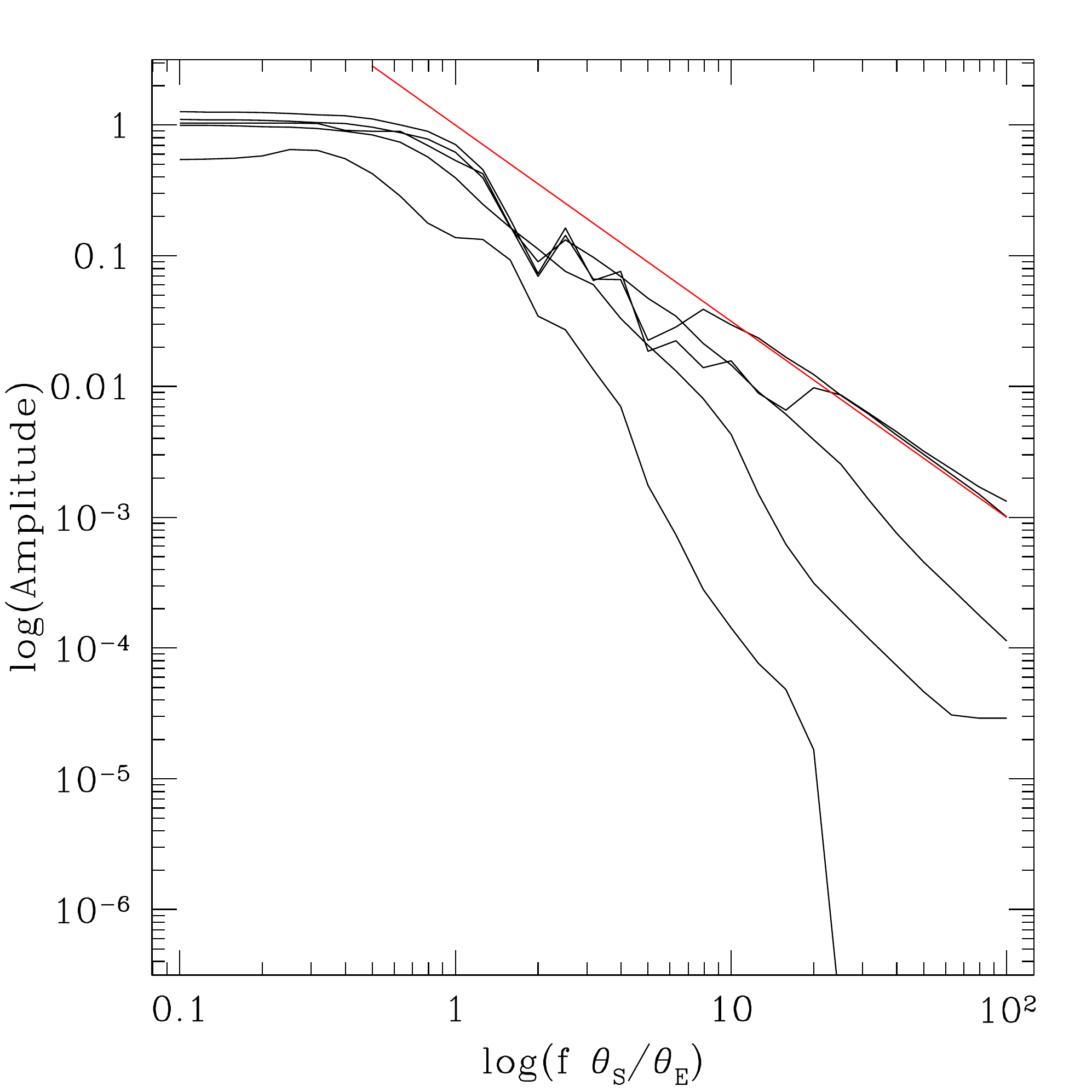}
\caption{The typical amplitude of the diffractive microlensing signal
  over the range $0.2 < v/R_E < 3$ as a function of the ratio of angular size
  of the source $\theta_s=R_s/D_s$ to that of the fringes.  The number
  of fringes across the source is given approximately by 
  $2/\pi f \theta_S/\theta_E$.  The red line shows
  the relation where the amplitude equals $(f
  \theta_s/\theta_E)^{-1.5}$.  From top to bottom
  $f=100,10^{1.5},10,10^{1/2}$ and 1.}
\label{fig:ampl}
\end{figure}

According to Fig.~\ref{fig:ampl}, the relative amplitude of the
variation is approximately equal to $(f \theta_s/\theta_E)^{-1.5}$
for large values of $f$ and $(f\theta_s/\theta_E)$.  The amplitude of
the power-law relation depends on what portion of the light curve is
examined; however, the exponent comes simply from the integral of a
sinusoid (Eq.~\ref{eq:4}) over the circular source 
\begin{equation}
\int_{-1}^1 d x \sqrt{1-x^2} \cos \left (k x + a\right ) = \frac{\pi \cos
  a }{k} J_1(k) \propto k^{-3/2} 
  \label{eq:8}
\end{equation}
where $k\propto f \theta_S/\theta_E$.  The final proportionality holds
for values of $k$ much greater than one.  The derivation of this
expression clearly only makes sense if $f\gg 1$ and $\theta_S \ll
\theta_E$

More generally the relative amplitude of the variation can be taken to
be a power law to give 
\begin{eqnarray}
\Delta f_\nu &\approx& A \left ( f \frac{\theta_s}{\theta_E} \right
)^{-\alpha} f_\nu = A f_{\nu,d}^{\alpha/2} f_\nu^{(2-\alpha)/2} \\
&=& f_{\nu,d}^{3/4} f_\nu^{1/4}
  \label{eq:9}
\end{eqnarray}
for $f_\nu > f_{\nu,d}$ and $f \gg 1$ and 
where
\begin{equation}
f_{\nu,d} = \frac{k T_b}{2\pi R_E^2} = 0.4 \mathrm{nJy}
\frac{T_b}{1000~\mathrm{K}} \frac{M_\oplus}{M_d} \frac{1~\mathrm{pc}}{D_d}
  \label{eq:10}
\end{equation}
and $T_b$ is the brightness temperature of the source and $f_\nu$ is the
flux density of the source.


\subsection{Detection}
\label{sec:detection}

For simplicity it can be assumed that the diffractive microlensing
event introduces a sinusoidal variation on the radio emission from the
source.  Although this variation is not strictly sinusoidal, it is
approximately sinusoidal in the limit of large $f$.  The Square
Kilometre Array is planned to have a sensitivity of $10^4 \mathrm{m}^2
\mathrm{K}^{-1}$ \citep{Schi07} or a noise level of $0.27$~Jy per
Nyquist sample.  Using the spectral properties of Gaussian noise, one
could detect a sinusoidal variation in the flux of the source down to
an amplitude of
\begin{equation}
\Delta f_\nu = 20.~\mathrm{nJy} \left [
1+0.21 \ln(N_\mathrm{Trials}) \right ]^{1/2}
\left (  \frac{B}{10 \mathrm{GHz}} \frac{t}{2 \mathrm{days}} \right
)^{-1/2}
\label{eq:12}
\end{equation}
with at most a one-percent false-positive rate.  The quantity $B$ is
the bandwidth of the observation and $t$ is its duration that is set
by the length of the typical microlensing event as well as observing
efficiencies.  In order to use a large bandwidth to decrease the noise
level, one must divide the large bandwidth up finely because the flux
varies at different rates at different observing frequencies.  Without
dividing the bandwidth the variation will be smeared out.  This is
similar in spirit to the process of dedispersion in pulsar searches
where the pulse phase depends on the observing frequency
\citep{1977puls.book.....M}.  Here the frequency of the variation
itself depends on the observing frequency.

This detection threshold does not include noise associated with the
source, either internal or external (such as scintillation).
Typically because the sources are not much brighter than the detection
threshold, the relative variation due to microlensing is large, so the
other sources of noise proportional to the strength of the source are
assumed to lie below the detection threshold.

\section{The Sources}
\label{sec:sources}

Although this detection limit is quite faint, it is unusual for
typical microlensing sources to be bright radio sources, so the effect
will be difficult to detect.  The sources can be divided into
essentially two classes.  The first are experiencing a microlensing
event due an intervening lens detected in the optical.  The second are
bright compact radio sources that could be lensed by nearby objects
serendipitously.  Because the Square Kilometre Array in principle can
look at all of the sources in a portion of sky all the time,
monitoring all of these sources is a possibility only limited by the
available computing power.

Because the current slew of radio telescopes are not sensitive enough
to detect typical sources for microlensing in the Galactic bulge and
the Magellanic clouds, observations of local analogues must provide
an estimate of the expected radio spectra of these objects.  

\subsection{Low-Mass Stars}
\label{sec:low-mass-stars}

To survey the types of common radio sources in the bulge, it is
natural to focus on local low-mass stars that also happen to be radio
sources.  Typically isolated low-mass stars do not exhibit strong
continuous radio emission.  Even at the modest distance of one parsec,
the radio emission from the quiet sun is only about one~$\mu$Jy
\citep{GalacticRadioAstronomy}.  At one kiloparsec this is down to
one~pJy, well below the detection threshold of even the SKA.
Fortunately, evolved low-mass stars exhibit much greater quiescent
radio emission than the sun; these objects will be the focus.

\subsubsection{Giants}
\label{sec:giants}

Arcturus ($\alpha$ B\"ootis) at 11.25 pc is one of the closest giant
stars to Earth
\citep{1997A&A...323L..49P}. \citet{1986AJ.....91..602D} presented
observations of this star at 2~cm and 6~cm with flux densities of
0.68~mJy and 0.28~mJy respectively.  At both wavelengths the emission
is well characterised by a brightness temperature of about $1.3\times
10^4$~K --- the source is larger at longer wavelengths.  At a distance
of 1~kpc the flux density would be
\begin{equation}
f_\nu \approx 24.~\mathrm{nJy} \nu_\mathrm{GHz}^{0.8} \left (
  \frac{1~\mathrm{kpc}}{D_s} \right )^{2}.
\label{eq:28}
\end{equation}
using the spectral index found by \citet{1986AJ.....91..602D}.  This
yields an expected amplitude variation of 
\begin{equation}
\Delta f_{\nu} \approx 12.~\mathrm{nJy} \left (\frac{M_\oplus}{M_d}
  \frac{1~\mathrm{pc}}{D_d} \right )^{3/4} \left (
  \frac{1~\mathrm{kpc}}{D_s} \right )^{1/2}.
\label{eq:29}\end{equation}
at 10~GHz.

\subsubsection{Asymptotic Giants}
\label{sec:asymptotic-giants}

The nearby variable M7-giant star Mira (o Ceti) provides an exemplar
for the rarer asymptotic giant stars observed by OGLE in the bulge of
our galaxy.  \citet{1997ApJ...476..327R} find that the radio emission
of Mira is well fit by a blackbody emission with $S_\nu \approx
6\nu_\mathrm{GHz}^2 \mu\mathrm{Jy} $, and Hipparcos found its distance
to be about 110~pc \citep{1997A&A...323L..49P}, yielding a diameter of
about 8~AU.  The signal of a diffractive lensing event on a Mira-like
star ($T_b=1500$~K) at 10~GHz is given by
\begin{equation}
\Delta f_{\nu} \approx 6.3~\mathrm{nJy} \left (\frac{M_\oplus}{M_d}
  \frac{1~\mathrm{pc}}{D_d} \right )^{3/4} \left (
  \frac{1~\mathrm{kpc}}{D_s} \right )^{1/2}.
\label{eq:22}
\end{equation}
Although an AGB star is typically more luminous that a giant both in
the optical and the radio,  flux only plays a subordinate role in the
expected diffractive microlensing variation, the significantly lower
brightness temperature of the AGB stars in the radio reduces the
expected signal.

\subsection{Supergiants}
\label{sec:supergiants}

The M2-supergiant Betelgeuse ($\alpha$ Orionis) provides an exemplar
for the red supergiant stars observed by OGLE in the Magellanic
clouds, assuming a distance of 48~kpc to the Large Magellanic Cloud.
\citet{1982ApJ...263L..85N} find that the spectrum of Betelgeuse is
well characterised by $S_\nu \approx 240 \nu_\mathrm{GHz}^{1.32}
\mu\mathrm{Jy} $, and \citet{2008AJ....135.1430H} give a distance of
197~pc to this object.  \citet{1998Natur.392..575L} found that the
photosphere at a wavelength of 7~mm subtends an ellipse about 95~mas
by 80~mas.
%
The signal of a diffractive lensing event on a Betelgeuse-like star
($T_b=2500$~K) at 10~GHz is given by
\begin{equation}
\Delta f_{\nu} \approx 3.0~\mathrm{nJy} \left (\frac{M_\oplus}{M_d}
  \frac{1~\mathrm{pc}}{D_d} \right )^{3/4} \left (
  \frac{48~\mathrm{kpc}}{D_s} \right )^{1/2}.
\label{eq:23}
\end{equation}
The increased luminosity of the supergiant Betelgeuse is offset by its
larger size and flatter spectrum.

\citet{1989AJ.....98.1831D} found that the B8-supergiant Rigel
($\beta$ Orionis) has a flux density of 270$\mu$Jy at 6.3~cm and a
brightness temperature of about 15000~K.  They argue for a spectrum
$f_\nu \propto \nu^{0.6}$ yielding the signal of a diffractive lensing
event on a Rigel-like star at 10~GHz of
\begin{equation}
\Delta f_{\nu} \approx 6.8~\mathrm{nJy} \left (\frac{M_\oplus}{M_d}
  \frac{1~\mathrm{pc}}{D_d} \right )^{3/4} \left (
  \frac{48~\mathrm{kpc}}{D_s} \right )^{1/2}.
\label{eq:24}
\end{equation}
assuming that the distance to Rigel is 237~pc
\citep{1997A&A...323L..49P}.  Even bluer stars yield even higher
variable flux densities.  Using the O4-supergiant HD~190429 as an
exemplar yields
\begin{equation}
\Delta f_{\nu} \approx 37.~\mathrm{nJy} \left (\frac{M_\oplus}{M_d}
  \frac{1~\mathrm{pc}}{D_d} \right )^{3/4} \left (
  \frac{48~\mathrm{kpc}}{D_s} \right )^{1/2}
\label{eq:25}   
\end{equation}
so a nearby diffractive lensing event would be detectable
against an early supergiant in the Large Magellanic Cloud (LMC) or
Small Magellanic Cloud (SMC).

For all of these estimates the only source of noise included is at the
receiver.  The radio emission of giants and supergiants may be
sufficiently noisy to increase the detection limits outlined in
\S~\ref{sec:detection}.

\subsection{Radio-loud Quasars}
\label{sec:radio-loud-quasars}

Nearby radio-loud quasars typically have brightness temperatures of about
$10^{11}$~K and flux densities of about 1~Jy, much brighter than
typical stars; however, they, of course, are much rarer and it is
less likely that nearby substellar objects will cause lensing events.
The assumption that the quasar lies at a distance of about 1~Gpc
and the lens lies in the solar neighbourhood
\begin{equation}
f_{\nu,d} = \frac{k T_b}{2\pi R_E^2} = 40 \mathrm{mJy}
\frac{T_b}{10^{11}~\mathrm{K}} \frac{M_\oplus}{M_d} \frac{1~\mathrm{pc}}{D_d}.
\label{eq:26}
\end{equation}
Any quasar with $f_\nu < f_{\nu,d}$ would exhibit large oscillations
in its flux during a diffractive microlensing event; whereas 
brighter sources would exhibit variations of 
\begin{equation}
\Delta f_\nu = 90. \mathrm{mJy} 
\left ( \frac{T_b}{10^{11}~\mathrm{K}} \frac{M_\oplus}{M_d}
  \frac{1~\mathrm{pc}}{D_d} \right )^{3/4} \left (
  \frac{f_\nu}{1~\mathrm{Jy}} \right )^{1/4}.
\label{eq:27}
\end{equation}
Although lensing events with quasars are likely to be rare, with the
SKA it may be possible to monitor a sample of bright quasars
continually and provide constraints on the number of freely floating
planets.   On the other hand radio emission of quasars is inherently
noisy.  This might make detecting the diffractive lensing events
difficult or produce false detections.   Either of these possibilities
merit further study.

\subsection{OGLE-II Sources}
\label{sec:ogle-ii-sources}

\begin{figure}
\includegraphics[width=3.4in]{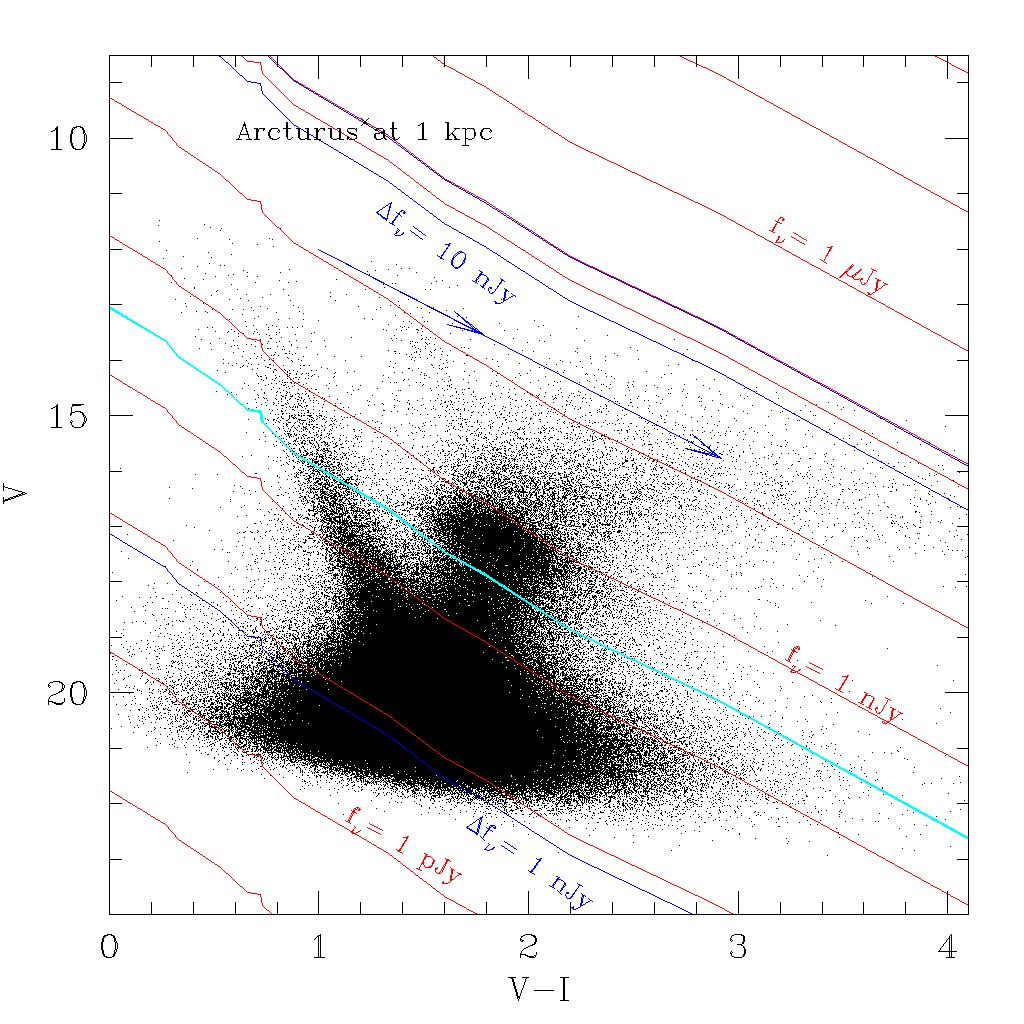}
\caption{The estimate of the radio flux density (red) and amplitude of
  its oscillation (blue) at 10~GHz for stars in the bulge field 1 of the
  OGLE-II survey \citep{2005AcA....55...43S, 1997AcA....47..319U}.
  The fluxes are normalaized using the values for Arcturus at 1~kpc
  Eqs.~\ref{eq:28} and~\ref{eq:29} (purple curve).  The cyan line
  denotes the value of $f_{\nu,d}$ (Eq.~\ref{eq:10}) for
  $T_b=13000$~K.  The minimum and maximum mean reddening lines derived
  from OGLE-II are given by the blue arrows
  \citep{2004MNRAS.349..193S}.}
\label{fig:ogle}
\end{figure}

To estimate the number of sources that could be detected in the radio
with the SKA and whose oscillatory flux could be detected, a detection
threshold of 10~nJy is chosen.  The radio flux density from each of the
OGLE-II sources also has to be modelled.   The relationship between
bolometric correction, $V-I$-colour and effective temperature given in
\citet{2007gitu.book.....S} provides a estimate of the angular size of
the stellar photosphere as a function of $V-I$ and $V$.  This
relationship depends only weakly on the luminosity class of the
stars.  The radio flux is assumed to be proportional to the angular
size of the photosphere and normalised to the values for Arcturus in
Eq.~\ref{eq:28} at 10~GHz.  These curves are shown in red in
Fig.~\ref{fig:ogle} and the corresponding values of $\Delta f_\nu$ are
shown in blue.   The normalisation to Arcturus is given in purple and
the value of $f_{\nu,d}$ in cyan.

The reddening in the OGLE-II fields is extensive
\citep{2004MNRAS.349..193S}.  The minimum and maximum mean reddenings
are given by the blue arrows, and they are nearly parallel with the
curves of constant radio flux, so the individual stars do not need to
be dereddened to get a rough estimate of their radio fluxes.  The
OGLE-II photometry database \citep{2005AcA....55...43S,
  1997AcA....47..319U} contains about 21000 OGLE-II bulge sources with
$\Delta f_\nu > 10$~nJy at 10 GHz and 85000 OGLE-II bulge sources with
$f_\nu > 10$~nJy at 10 GHz.  This should be compared with the number
of red clump, red giants and red supergiants that
\citet{2006ApJ...636..240S} used to detect microlensing events
(1,084,267 stars); therefore, about eight percent of microlensing events
may be detectable in the radio.  Furthermore, if the lens is about an
Earth mass within a few parsecs about two percent of the sources would
yield a detectable oscillatory signal.  Only a handful of stars in
the OGLE-II LMC and SMC fields are sufficiently bright to yield a
detectable radio source.

Of course all of these estimates assume that the sources only emit
essentially thermal radiation from near their optical photospheres.
In fact some fraction of the stars in the bulge and Magellanic Clouds
will be radio loud, but it will take an instrument with a sensitivity
rivalling that of the SKA to determine their number; therefore, these
estimates provide a moderately conservative lower limit on the
number of expected radio sources in the bulge at this flux density.

The optical depth of a particular set of objects to microlensing is
proportional to their mass fraction; the area subtended by the
Einstein radius of a particular lens is proportional to its mass, so
the total area (and therefore optical depth) of a particular group of
the lenses is proportional to their total mass.  A simple model is to
assume that the density of lenses is constant out to some maximal
distance, yielding
\begin{equation}
\tau = \frac{2\pi G}{c^2} \rho l^2 = 2.3 \times 10^{-8} \left (
  \frac{l_\mathrm{max}}{1~\mathrm{kpc}} \right )^2 \frac{\rho_\mathrm{lens}}{\rho_\mathrm{Hipparcos}}
\end{equation}
where the dynamical mass estimate of the local disk ($0.076 \msun
\mathrm{pc}^{-3}$) from Hipparcos \citep{1998A&A...329..920C} provides
an upper limit on the mass density of lenses within the Galactic disk.
This is of course a rough model of the lensing optical depth as the
density of lenses will change along the line of sight in a realistic
Galactic model, and some substellar lenses may lie in the halo.
However, only a rough estimate is warranted because the density of
substellar objects is unknown and it is unlikely to exceed about
one-tenth of the stellar density \citep{2002ApJ...567..304C}.  This
yields a rough upper limit on the lensing optical depth of $\tau
\lesssim 2 \times 10^{-9}$.  If one monitors the $10^5$ stars in the
OGLE-II survey that are also detectable with the SKA and assumes an
event duration of one day, one would expect to find a single
diffractive lensing event during which the source and lens align to
within one Einstein radius about 14~years.  Searching for lensing
events among fainter stars will not improve this situation because
they would probably not be detectable with the SKA.  However,
monitoring a larger field of stars (going wide rather than deep) would
increase this rate proportionally to the number of bright stars.

\section{Conclusions}
\label{sec:conclusions}

The Square-Kilometer Array will offer an unprecedented view of the
radio sky both in terms of sensitivity and cadence.  These two factors
make it a potential instrument to search for microlensing in the
radio.  In the spectral range of the SKA diffraction is important to
understand the microlensing by objects less massive than Jupiter.  The
diffractive elements of the microlensing event offer a
model-independent way to estimate the mass of the lens that can only
otherwise be inferred through statistical arguments.

As \citet{1991ApJ...374L...5P} argue quasars are among best the best
candidates for finding this effect.  However, bright quasars are
relatively rare and possibly their inherent noise is sufficient to
mask the lensing-induced oscillations.  This letter argues that the
oscillations are also detectable for blue supergiants in the LMC and
nearly detectable for giants in the bulge. However, the likelihood of
a lensing event against the former is very small due to the small
number of blue supergiants in the Magellanic clouds.  Because of the
relative rarity of blue supergiants and quasars on the sky, most
promising potential sources for diffractive microlensing are giant
stars --- although even in this case the rate is quite low using the
OGLE-II stars.

Of course all of these estimates assume that the radio emission of
giant and supergiant stars is similar to the few local detections.  In
fact understanding the emission from these stars is one of the
objectives of the SKA, and if the emission turns out to be stronger
the prospects are even better. Even if the oscillations are not
observable, searching for the frequency at which the magnification
deviates from the geometric optics result would indicate that $f\sim
1$ and also provide a mass estimate.  Such a measurement would simply
require that the source be detectable with the SKA, a much less
stringent requirement than detecting the oscillations themselves.

Although this paper has outlined the potential benefits of observing
lensing events in the radio that derive from the diffractive aspects
of microlensing, there are several important reasons to follow up
lensing events in the radio even if the diffractive effects are
unlikely to be observable.  In particular the lenses themselves may be
radio sources --- this would yield an estimate of the lens proper
motion as well as its identity.  If the source can be detected in
the radio, observations in the radio could probe the astrometric
signature of the lensing event, providing additional constraints.

Several potential substellar lensing events have already been
observed, and over the next decades many more will be found as the
monitoring campaigns become more sensitive and extensive.  Following
these events up in the radio with an instrument like the SKA may
provide important constraints on their properties and the constituents
of the solar neighbourhood.

\section*{Acknowledgments}

The Natural Sciences and Engineering Research Council of Canada,
Canadian Foundation for Innovation and the British Columbia Knowledge
Development Fund supported this work.  This research has made use of
NASA's Astrophysics Data System Bibliographic Services and the OGLE
online database.

\bibliographystyle{mn2e}
\bibliography{mine,physics,math,oort,paper2}
\label{lastpage}
\end{document}